# Detailed Radiative Transport Modeling of a Radiative Divertor

A. S. Wan, H. E. Dalhed, H. A. Scott, D. E. Post,[a] T. D. Rognlien

Lawrence Livermore National Laboratory, Livermore, CA 94550, U.S.A.
[a]ITER Joint Central Team, San Diego Co-Center, San Diego, CA 92037, U.S.A.

## Abstract

An effective radiative divertor maximizes the utilization of atomic processes to spread out the energy deposition to the divertor chamber walls and to reduce the peak heat flux. Because the mixture of neutral atoms and ions in the divertor can be optically thick to a portion of radiated power, it is necessary to accurately model the magnitude and distribution of line radiation in this complex region. To assess their importance we calculate the effects of radiation transport using CRETIN, a multi-dimensional, non-local thermodynamic equilibrium simulation code that includes the atomic kinetics and radiative transport processes necessary to model the complex environment of a radiative divertor. We also include neutral transport to model radiation from recycling neutral atoms. This paper presents a case study of a high-recycling radiative divertor with a typical large neutral pressure at the divertor plate to estimate the impact of H line radiation on the overall power balance in the divertor region with consideration for line opacities and atomic kinetics.

## 1. Introduction

The peak heat load on the divertor plate of the next-generation tokamak, such as ITER, after adjusting for the effect of field line tilting, bremsstrahlung, and radiation loss from the central plasma, is expected to be of order 35–70 MW/m$^2$ [1] if very little of the divertor power is radiated to the walls. This high heat load will result in excessive temperatures in candidate divertor plate materials, such as Be, and significantly impact our ability to obtain a feasible engineering design. One possible concept to reduce the heat load on the divertor plate is to utilize atomic processes to spread out the energy deposition, or effectively increase the "wetted" area of energy deposition in the divertor chamber [1,2]. A large neutral pressure can build up due to the recycling of H ions neutralized at the divertor plate. Large radiation losses, from both continuum and line radiation, can spread the energy to the total surface area available in the divertor chamber. However, the mixture of the fuel and impurity neutral atoms and ions is optically thick to a portion of the radiated power [3,4]. Therefore it is necessary to accurately model the magnitude and distribution of both the line and continuum radiation in the radiative divertor.

The primary purpose of this paper is to present a sample calculation, using our non-local thermodynamics equilibrium (NLTE) modeling code, CRETIN, that illustrates how our model can be used to assess the importance of line and continuum radiation in the overall power balance of a high-recycling radiative divertor. We will address the collisional and radiative processes that are included in our atomic models which are important in the relevant regime in the divertor regions of both ITER and present-day tokamaks. We then use an one-dimensional (1-D) configuration derived by taking the parameters just outside the magnetic separatrix from a deuterium gas-puff shot in DIII-D [5] to illustrate the important radiation trapping effects of a radiative divertor.

**2. Code Description and Relevant Atomic Physics**

CRETIN is a multi-dimensional NLTE simulation code that includes the atomic kinetics and radiative transport processes necessary to model the complex environment of a radiative divertor. Atomic processes include electron-ion and ion-ion collisions, photo-ionization and photo-excitation, and Auger processes. Rates for inverse processes are done via detailed balance. Radiative processes include treatments of bound-bound, bound-free, and free-free processes.

CRETIN is a subset of the code GLF, which has been described in Ref. [6]. CRETIN does not include hydrodynamics, otherwise its simulation capabilities are virtually identical to GLF. In this study, plasma density and temperature profiles are fed into CRETIN, which then calculates distributions of ions, atoms, and radiation, for either time-dependent or steady-state conditions.

One further capability has been added to CRETIN to aid in divertor modeling: transport of neutral particles using a diffusion model [7]. Since the code does not transport ions, the diffusion is supplemented with the condition that the total mass density profile remains unchanged from the input profile. Transport of neutral particles can significantly affect the ionization balance near the divertor plate and is included in our analysis. The flux of the reentrant neutrals at the boundary is assumed to be proportional to the ion flux at the boundary.

The atomic models use the NLTE rate matrix approach [8], which consists of specifying the energy levels and the rates that connect these levels. The range of energy levels depends on the background plasma conditions and radiation fields. The complexity of the level structure likewise depends on the type of calculation desired. For detailed line radiation transport calculations, atomic configurations are described by multiconfiguration jj coupling [9], including configuration interaction using fully relativistic Dirac-Hartree-Fock self-consistent field wave functions [10]. This description is usually used for the principal shells involved in the radiative transitions of interest. The model is completed by specifying the levels of all remaining principal shells using a nonrelativistic Hartree-Fock-Slater jj-configuration average prescription [11,12], which may be further averaged by assuming statistical population distribution in the averaged manifolds.

Rates connecting the levels depend on the complexity of the level description. For the relativistic jj-coupled levels, bound-bound absorption oscillator strengths and state-to-state bound-free photoionization cross sections are determined using the multiconfiguration jj-coupled relativistic wave functions. Electron collisional bound-bound transitions are computed using a fully relativistic multiconfiguration distorted wave approach [13]. Electron collisional ionization rates are computed using a semi-relativistic distorted wave approach [14]. For energetically allowed transitions, fully relativistic multiconfiguration Auger rates are computed [15]. Dielectronic recombination and its inverse are also included [16]. For the averaged levels, various classical or semi-classical methods are used to determine the connecting rates [17,18].

## 3. Numerical Modeling

We present here a case study demonstrating the importance of including both line and continuum radiative processes in modeling a radiative divertor. We use 1-D electron and ion temperatures ($T_e$, $T_i$) and mass density ($\rho$) profiles similar to that of a deuterium gas-puffing shot in DIII-D [5] to approximate a high-recycling radiative divertor with substantial neutral pressure at the divertor plate. Total H (ions and neutrals) number densities, fixed by the input profile, range from $4 \times 10^{14}$ cm$^{-3}$ at the plate to $8 \times 10^{13}$ cm$^{-3}$ near the hot core plasma. $T_e$ at the divertor plate (x = 0) is 2 eV. The neutral diffusion is determined by charge exchange and ionization processes with the ionization mean free path for neutrals at these conditions of order of 50 cm. $T_e$ changes rapidly in a narrow ionization front (x = 15–17 cm, shaded region in Fig. 1) from 2 to 20 eV and reaches 40 eV at the core end of this problem (x = 40 cm).

We use CRETIN to calculate the steady-state condition, i.e., on time scales much longer than the dominant atomic rates. This NLTE condition is determined by detailed atomic processes, including all relevant collisional and radiative processes, mediated by photons, electrons, and ions, which determine the final ionization balance. The accuracy of our calculations depends on the choice of suitable atomic models. For this H-only case, we use a H model with excited neutral H levels averaged over the principle quantum number up to n = 10.

We compare the ionization balance and radiation energy density and power fluxes for three radiative transport models: (1) we calculate the Lyman-$\alpha$ radiation assuming the plasma is optically thin to the line; (2) full treatment of the Lyman-$\alpha$ line by detailed line transfer; and (3) detailed line transfer treatment with many H lines up to the $n = 6$ level.  We also transfer the continuum radiation by using a multiple group energy bin structure with fine energy bins up to 14 eV and a maximum energy of 600 eV.  The line transport enforces consistency between populations and line strengths through a complete linearization procedure.  For these calculations, the effects of Doppler shifting due to bulk plasma motion is negligible.

Figure 1 plots the spatial ion and neutral density profiles for cases (1) and (2).  The neutral density rapidly decays from the divertor plate due to ionization in the region between the ionization front and the divertor plate.  The drop in ion density from the ionization front toward the core is dictated by the input density profile.  Due to the large opacities of the H lines in this cold dense region, where the optical depth for the Lyman-$\alpha$ line for case (2) is ~ 30, the line radiation is trapped and excites the ground state population up to higher levels.  The excited electrons can then be collisionally ionized and thus change the overall ionization balance.  This effect is illustrated in Fig. 1 by directly comparing the neutral and ion density profiles of case (2), with the optically thick Lyman-$\alpha$ line, and case (1) without line opacity effects.

Figure 2 plots the spatial distributions of the directional line fluxes for the three cases studied here.  The solid lines represent fluxes radiating back to the core while the dashed lines represent fluxes toward the divertor plate.  In case (1) where we assume the line is optically thin, we can clearly see the line flux building up in the cold, high-neutral-density region between the ionization front and the divertor plate due to the strong line emissivity, but remaining flat over the hot region near the core where neutral density is very low.  As shown in Fig. 1, an optically thick Lyman-$\alpha$ line is sufficiently strong to perturb the ionization balance of the cold plasma near the plate.  The line also strongly affects the distribution of the excited populations of the neutral H, changing the emissivity of the plasma. In case (1), in addition to ignoring the attenuation due to line photon reabsorption, we ignore this change in emissivity as well. These are two important effects in an optically thick plasma which act in opposite senses, but they do not cancel. The net error in the fluxes is a factor of two in this case, represented in Fig. 2, and will increase with higher neutral densities and larger line opacities. When reabsorption is included in the line transfer treatment, the line flux radiating toward the plate is strongly attenuated by the large-opacity, high-neutral-fraction plasma near the plate. Note that even if the Lyman-$\alpha$ line is very optically thick, there is still a significant flux at the plate. This is because even though the large optical depth prevents photons from escaping in the line center, they can still effectively escape in the line wings.

Table 1 summarizes the magnitude of line fluxes exiting the two boundaries for the three cases and clearly demonstrates the need for detailed treatments of line transport for a radiative divertor in which the neutral density is sufficiently high for the mean free path, $\lambda \sim 0.2/n_o(10^{14}$ cm$^{-3})$ cm, of Lyman-$\alpha$ radiation to be of order of a few centimeters. Without accounting for reabsorption, fluxes at the plate and back to the core are roughly identical and are significantly higher than fluxes obtained with correct treatments of the lines. With the detailed treatments of line opacity and emissivity, we observe a larger fraction of the line flux directed back to the core, thus reducing the heat load on the divertor plate. We expect this asymmetric distribution of fluxes toward and away from the divertor plate would increase for larger neutral densities and higher line opacities. Extending the geometry to 2-D would show that the radiation is spread over the entire volume of the divertor with the chamber walls taking a majority of the radiating power and thereby significantly increase the total "wetted" area of the heat flux, thus reducing the total heat load on the divertor plate. Comparisons between cases (2) and (3) show that Lyman-$\alpha$ is the dominant line in a pure H plasma. In case (2) we observe a Lyman-$\alpha$ line flux of 0.146 MW/m$^2$ directed out to the core, accounting for close to 90% of all the H line fluxes in case (3).

The continuum flux is not strongly attenuated but we can ignore the contribution of the continuum flux since line fluxes dominate over the continuum with the intensities of the continuum fluxes for these cases of order of 0.004 MW/m$^2$. The continuum flux also increases when the line transport is included. This is due to the changes in ionization balance since the continuum radiation is mostly due to bremsstrahlung which is directly proportional to the product of $n_i n_e$.

Ionization and recombination rates depend sensitively not only on local values of $n_e$, $T_e$, and neutral density [1] but also on the detailed global emissivity and opacity of the radiation field. Under conditions in which the frequency dependent optical depth becomes appreciable, it is essential to model the ionization balance with detailed radiation transport using NLTE rate equations. Details of the spectrum in one region of the plasma affect the ionization balance in other regions. For example, Figure 3 shows the effect on ionization balance of Lyman-$\alpha$ line pumping for line intensities characteristic of those found in our typical divertor calculations. For a H plasma with $T_e = 1.5$ eV and $n_e = 10^{12}$/cm$^3$, the coronal model would predict an average of 0.44 bound electrons. The local intensity in the Lyman-$\alpha$ line will increase due either to trapping or to a nearby line emitting region, both of which are present in the radiative divertor. This line radiation will excite population from the n = 1 level to the n = 2 level, greatly increasing the collisional ionization rate. As can be seen, even modest departures of the radiation field from the coronal model can markedly change the ionization balance.

## 4. Summary and Future Research Directions

The design of an effective radiative divertor is on the critical path of the ITER EDA phase. This paper demonstrates the importance of the detailed radiative modeling in assessing the role of both line and continuum radiation losses in the design of a radiative divertor. We present here a case study of the impact of detailed radiation transport modeling on a 1-D, high-recycling, low-temperature divertor configuration using CRETIN, a NLTE simulation code that includes all the relevant atomic and radiative processes. The addition of neutral diffusion to CRETIN is important in the self-consistent modeling of the neutral penetration into the divertor plasma.

The high neutral-density plasma near the divertor plate is optically thick to the line radiation and results in reduced power fluxes to the divertor plate. Ignoring the reabsorption of the line radiation results in significant over-estimates of the H radiation fluxes to the divertor plate and to the rest of the divertor. The H line radiation is also intense enough to alter the ionization balance of the mostly neutral plasma near the divertor plate. If the line radiation strongly changes the overall energy and ionization balances, we may need to couple the detailed radiative modeling calculations to multi-dimensional modeling codes such as UEDGE [7]. Such calculations will be computationally expensive, but they may be the only valid way to produce a self-consistent simulation of the radiative divertor.

Recent experiments on DIII-D [19] and JT-60U [20] clearly illustrate the importance of impurity radiation on the overall energy balance of a divertor plasma. In this paper we focused on a pure H plasma. The likely candidate material for the ITER divertor plate will be Be. In the future we plan to address the impact of a low-impurity (~5%) plasma on the overall energy balance, which needs to take account of all possible Be lines that may contribute to the energy balance.

**Acknowledgment**

This work is performed under the auspices of the U. S. DOE by LLNL under the contract number W-7405-ENG-48 and is partially supported by the ITER US Home Team.

**Table Captions**

Table 1.  Line fluxes (in MW/m$^2$) exiting the problem boundaries (to divertor plate or back to core) for the three case studies.

|  | back to core | to divertor plate |
|---|---|---|
| Lyman-$\alpha$ w/o line opacity | 0.428 | 0.427 |
| detailed line transfer of Lyman-$\alpha$ | 0.146 | 0.071 |
| detailed line transfer of many H lines | 0.164 | 0.091 |

Table 1.   Wan, Dalhed, Scott, Post, Rognlien

**Figure Captions**

Figure 1.  Spatial profiles of the neutral and ion densities for case (2), in solid lines, and case (1), in dashed lines. The inclusion of a strong, optically thick Lyman-$\alpha$ line of case (2) increases the ion population due to the combined processes of first excitation of the ground state electrons and subsequent collisional ionization.

Figure 2.  Spatial profiles of the directional line fluxes for the three radiation models. Solid lines correspond to fluxes toward the core and dashed lines correspond to fluxes toward the divertor plate.

Figure 3.  The effect on ionization balance of Lyman-$\alpha$ line pumping for line intensities characteristic of those found in our typical divertor calculations for a H plasma with $T_e$ = 1.5 eV and $n_e = 10^{12}/cm^3$.

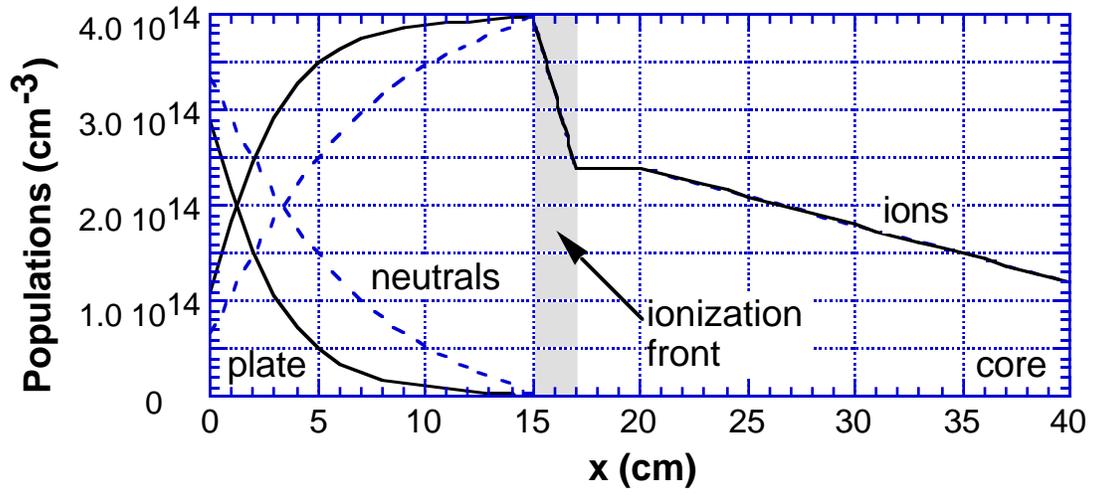

Figure 1.  Wan, Dalhed, Scott, Post, Rognlien

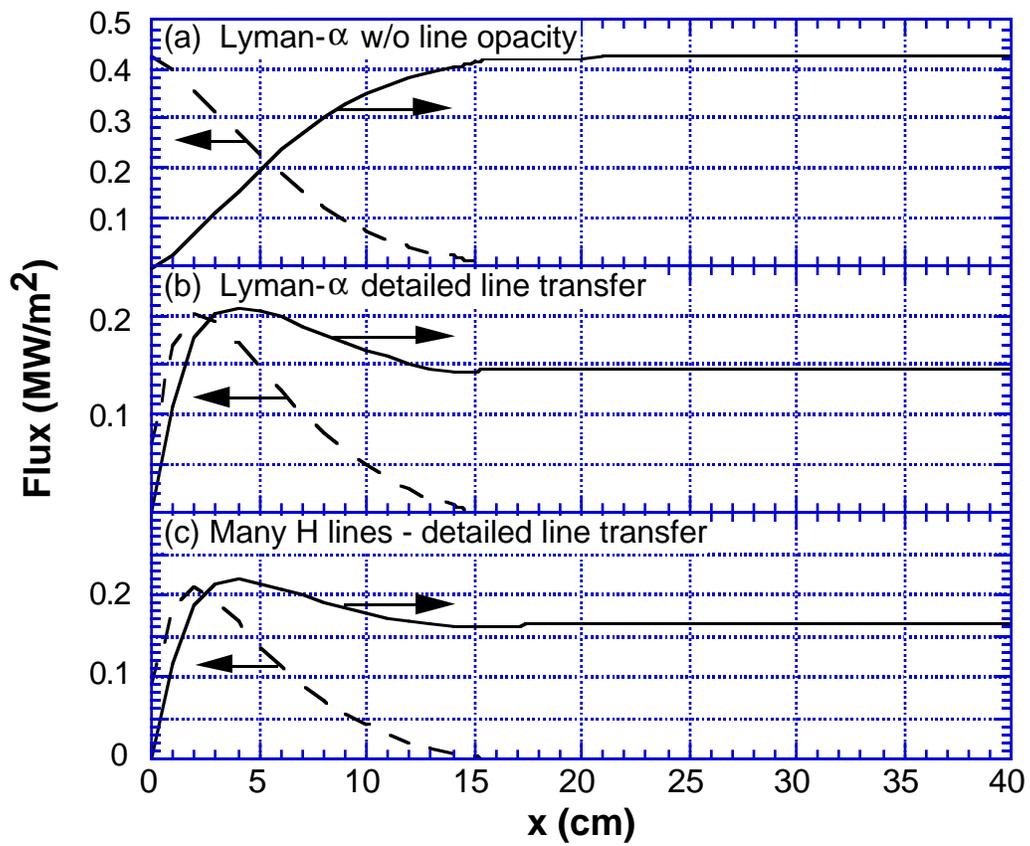

Figure 2. Wan, Dalhed, Scott, Post, Rognlien

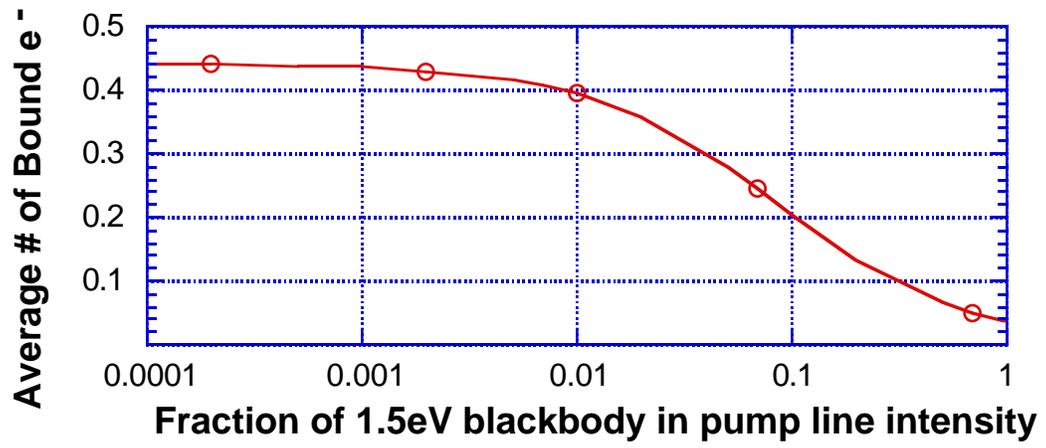

Figure 3.   Wan, Dalhed, Scott, Post, Rognlien